\documentstyle[12pt,epsf,bezier,axodraw]{article}
\begin{document}
\setlength{\baselineskip}{0.33 in}
\catcode`@=11
\long\def\@caption#1[#2]#3{\par\addcontentsline{\csname
  ext@#1\endcsname}{#1}{\protect\numberline{\csname
  the#1\endcsname}{\ignorespaces #2}}\begingroup
    \small
    \@parboxrestore
    \@makecaption{\csname fnum@#1\endcsname}{\ignorespaces #3}\par
  \endgroup}
\catcode`@=12
\newcommand{\newc}{\newcommand}
\newc{\gsim}{\lower.7ex\hbox{$\;\stackrel{\textstyle>}{\sim}\;$}}
\newc{\lsim}{\lower.7ex\hbox{$\;\stackrel{\textstyle<}{\sim}\;$}}
\def\NPB#1#2#3{Nucl. Phys. {\bf B#1} #3 (19#2)}
\def\PLB#1#2#3{Phys. Lett. {\bf B#1} #3 (19#2)}
\def\PRD#1#2#3{Phys. Rev. {\bf D#1} #3 (19#2)}
\def\PRB#1#2#3{Phys. Rev. {\bf B#1} #3 (19#2)}
\def\PRL#1#2#3{Phys. Rev. Lett. {\bf#1} #3 (19#2)}
\def\PRT#1#2#3{Phys. Rep. {\bf#1} #3 (19#2)}
\def\MODA#1#2#3{Mod. Phys. Lett. {\bf A#1} #3 (19#2) }
\def\ZPC#1#2#3{Zeit. f\"ur Physik {\bf C#1} #3 (19#2) }
\def\ZPA#1#2#3{Zeit. f\"ur Physik {\bf A#1} #3 (19#2) }
\def\bdm{\begin{equation}}
\def\edm{\end{equation}}
\def\bea{\begin{eqnarray}}
\def\eea{\end{eqnarray}}
\def\phift{\phi^4_2}
\def\sp{symmetric phase}
\def\bsp{broken symmetry phase}
\def\cc{coupling constant }
\def\EQFT{Euclidean Quantum Field Theory}
\def\bm2{\mu_0^2}
\def\bm2l{\mu_{0_\ell}^2}
\def\laml{\lambda_\ell}
\def\rc{renormalization condition}
\def\rcz{renormalization constant}
\def\ia{\tau_{\rm INT}}
\vsize 8.7in
\def\singlespace{\baselineskip 11.38 pt}
\def\halfagainspace{\baselineskip 17.07 pt}
\def\doublespace{\baselineskip 22.76 pt}
\def\medspace{\baselineskip 17.07 pt}
\font\headings=cmbx10 scaled 1200
\font\title=cmbx10 scaled 1200
\halfagainspace

\begin{titlepage}
\begin{flushright}
{\large
VPI--IPPAP--97--9 \\
hep-lat/9712008\\
December 3, 1997 \\
}
\end{flushright}
\vskip 2cm
\begin{center}
{\Large {\bf Monte Carlo Simulation Calculation of Critical Coupling Constant for Continuum
 $\phift$} }
\vskip 1cm
{\Large 
Will Loinaz\footnote{E-mail: {\tt loinaz@vt.edu}}$^{a}$,
R.S. Willey\footnote{E-mail: {\tt willey@vms.cis.pitt.edu}}$^{b}$\\}
\vskip 2pt
$^{a}${\large\it Institute for Particle Physics and Astrophysics,\\
 Physics Department, Virginia Tech, Blacksburg, VA 24061-0435, USA}\\
\smallskip
$^{b}${\large\it Department of Physics and Astronomy\\
 University of Pittsburgh, Pittsburgh, PA 15260, USA}\\
\end{center}

\vspace*{.3in}
\begin{abstract}
We perform 
a Monte Carlo simulation calculation of the critical coupling constant
for the continuum ${\lambda \over 4} \phift$ theory.  
The critical coupling constant
we obtain is $\left[{\lambda \over {\mu^2}}\right]_{crit}=10.24(3)$.
\end{abstract}
\end{titlepage}
\setcounter{footnote}{0}
\setcounter{page}{2}
\setcounter{section}{0}
\setcounter{subsection}{0}
\setcounter{subsubsection}{0}
\setcounter{equation}{0}

\newpage
\section{\bf Introduction}
The $\phift$ field theory, specified by the (Euclidean) Lagrangian
\bdm
 {\cal L}_E = \frac{1}{2}(\nabla\phi)^2+\frac{1}{2}\mu_0^2\phi^2 
+\frac{\lambda}{4}\phi^4  \label{Leuc}
\edm
has solutions in a symmetric phase in which the discrete symmetry of the 
Lagrangian, $\phi \rightarrow -\phi$, is manifest, i.e. $<\phi>=0$ and 
there is no trilinear coupling. It also has solutions in a broken 
symmetry phase with $<\phi>\neq 0$ and induced trilinear couplings 
proportional to $<\phi>$. 

There exist both elegant heueristic \cite{chang} and rigorous \cite{gj}
mathematical proofs of the existence of this phase structure, but there is 
no rigorous result for the critical value of some coupling constant which
separates these two phases. There do exist numerous approximate calculations, 
and in the case of the lattice model (lattice spacing $a>0$), the critical 
line in the $\mu_0^2,\lambda$ plane which separates the two phases is 
known to some numerical accuracy by Monte Carlo simulation.

For the \EQFT (EQFT), which is the continuum limit ($a\rightarrow 0$) of the 
lattice model, the first step is to specify what finite dimensionless 
\cc is to be used to parametrize the solutions and for which we are to 
determine the critical value.

For the lattice model, these considerations are straightforward. There 
are two parameters, $\mu_0^2,\lambda$ in ${\cal L}_E$, and there is the
lattice spacing $a$. In $d=2$ both $\mu_0^2$ and $\lambda$
 have dimension mass squared. (We 
assume the infinite volume limit, $L\rightarrow \infty$). So there are two 
independent dimensionless parameters which may be taken to be the two
Lagrangian parameters measured in units of inverse lattice spacing 
squared,
\bdm
   \lambda_\ell = \lambda a^2, \;\;\;\; \mu_{0_\ell}^2 = \mu_0^2 a^2 
      \label{params}.
\edm
In this parametrization,the phase diagram of the lattice model consists 
of the critical line in the $\bm2l,\laml$ plane, determined by Monte Carlo
simulation (Fig. 1).

\begin{figure}
\centering
\epsfysize=2in  
\hspace*{0in}
\epsffile{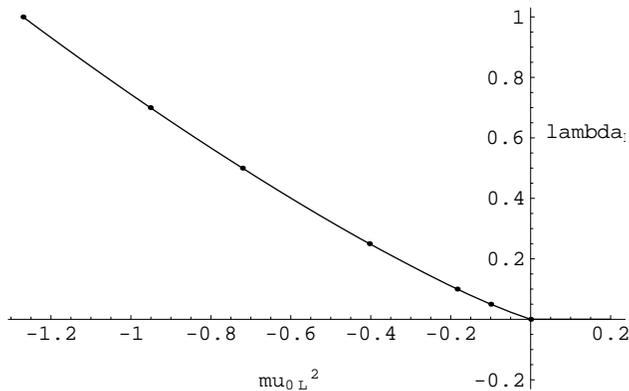} 
\caption{Phase Transition Line in the $\lambda_l, \mu^2_{0l}$ Plane}
\label{critfig1_ray} 
\end{figure}

The situation for the EQFT is more complicated. There is an infinite mass 
renormalization. The bare mass parameter has to be tuned to infinity as 
the continuum limit is taken, $\mu_0^2 \sim \mu^2\;\ln{\frac{1}{a}}$, where 
$\mu^2 $ is some finite renormalized mass squared. Thus 
$\mu_0^2$ cannot be used 
as a parameter of the continuum solution. In $d=2$, the field strength and 
\cc renormalizations ($Z_{\phi},Z_{\lambda}$) are finite and can be     
disregarded in the study of the phase structure of the theory. Furthermore,
 since the  dimensionful coupling constant  $\lambda$ is independent of $a$, 
and 
$\mu_0^2$ diverges only logarithmically with $a$, both $\laml,\bm2l$ go to zero
in the continuum limit, $a \rightarrow 0$. That is, the EQFT limit is the 
single point at the origin of the $\bm2l,\laml$ plane in  Fig. 
\ref{critfig1_ray}. Taking the 
limit $a \rightarrow 0$ reduces the number of of independent dimensionless 
parameters from two to one.

     The required mass renormalization can be written as a simple 
reparametrization of ${\cal L}_E$ (\ref{Leuc}). Let
\bdm
      \mu_0^2 = \mu^2 - \delta\mu^2.   \label{dmu}
\edm
Then
\bdm
  {\cal L}_E = \frac{1}{2}(\nabla\phi)^2+\frac{1}{2}\mu^2\phi^2 + 
  \frac{\lambda}{4}\phi^4 -\frac{1}{2}\delta\mu^2 \phi^2  \label{Lren}
\edm 
There is still a substantial freedom of choice in the definition of
the finite renormalized mass parameter, $\mu^2$. The ultraviolet, 
$\ln{\frac{1}{a}}$, dependence of $\mu_0^2$ is moved entirely to the counter 
term $\delta\mu^2$; but the separation of the finite part of $\mu_0^2$ into
$\mu^2$ and $\delta\mu^2$ is only determined when a renormalization 
condition is specified. The dimensionless effective \cc $f_{\mu^2}=\frac 
{\lambda}{\mu^2}$ then
manifestly depends on the choice of renormalization condition which fixes 
the finite part of $\delta\mu^2$. For example, we could take $\mu^2 = m_*^2$,
the pole mass, by choice of renormalization condition
\bdm
  0 = G^{-1}(p^2= - m_*^2)    \label{mrc1}
\edm
and dimensionless coupling
\bdm
   g = \frac{\lambda}{m_*^2}   \label{g}
\edm
A closely related choice, convenient for lattice Monte Carlo measurement, 
is to take $\mu^2 = m^{\prime 2}$, defined by the renormalization condition
\bdm
  m^{\prime 2} = G^{-1}(p^2 = 0)   \label{mrc2}
\edm
and
\bdm
      g^{\prime} = \frac{\lambda}{m^{\prime 2}}.   \label{gp}
\edm
In fact, neither of these choices provide a dimensionless \cc whose value 
distinguishes between the two phases of the theory - because the 
renormalization conditions themselves do not distinguish between the two 
phases. Either \rc, (\ref{mrc1}) or (\ref{mrc2}), 
can be implemented perturbatively 
in either the symmetric phase or the broken symmetry phase. 
The phase has to be specified by ansatz,
$<\phi> = 0$ or $<\phi>\neq 0$, so that one can perturb about the correct 
stable vacuum. Then $g$ or $g^{\prime}$ can take on arbitrarily 
small values in either phase.  

A dimensionless \cc whose critical 
value separates the two phases is provided by choosing the mass       
renormalization to be equivalent to normal ordering the interaction in 
the interaction picture in the symmetric phase. From (\ref{Leuc}),(\ref{dmu}) 
\bdm
    G^{-1}(p^2) = p^2 + \mu_0^2 +\Sigma_0(p^2) = p^2 +\mu^2 +\Sigma(p^2) 
      \label{sigma}
\edm
and for $\mu^2 > 0$
\bdm
     \Sigma(p^2) = 3\;\lambda\; A_{\mu^2}-\delta\mu^2+\mbox{two-loop}.
         \label{sigma2} 
\edm
      $A_{\mu^2}$ in the continuum limit is the ultraviolet divergent 
Feynman integral
\bdm
   A_{\mu^2}=\int\;\frac{d^2p}{(2 \pi)^2}\frac{1}{p^2+\mu^2}.  \label{A} 
\edm
On the lattice,
\bdm
   A_{\mu^2} = \frac{1}{N^2}\sum_{k_{1}=1}^{N}\sum_{k_{2}=1}^{N}\;\frac{1}{
   \mu_{\ell}^2 + 4\;\sin^2\;\frac{\pi k_{1}}{N}+ 4\;\sin^2\;\frac{\pi k_{2}}{N}    } 
    \label{A2} 
\edm 
The `leaf' diagram (Fig. 2) 
which gives the integral $A_{\mu^2}$ is the only 
divergent Feynman diagram of the theory in $d=2$.

\begin{center}\begin{picture}(300,100)(0,0)

\Line(25,50)(125,50)
\CArc(75,70)(20,0,360)

\end{picture} \\ {\hskip 10 pt Figure 2:  Leaf Diagram}
\end{center}

 Thus the \rc
\bdm
   \delta\mu^2 = 3\;\lambda\;A_{\mu^2}    \label{mrc3} 
\edm
   removes all ultraviolet divergence from the perturbation series based 
on the renormalized parametrization given by (\ref{Lren}) and (\ref{mrc3}).
\bea
  {\cal L}_E & = & \frac{1}{2}(\nabla\phi)^2 + \frac{1}{2}\mu^2\phi^2 + 
                   \frac{\lambda}{4}\phi^4 -\frac{3}{2}\lambda\;A_{\mu^2}
                   \phi^2  \nonumber   \\
            & = & \frac{1}{2}(\nabla\phi)^2 + \frac{1}{2}\mu^2\phi^2 + 
                  \frac{\lambda}{4} :\phi^4:_{\mu^2}   \label{LNO}
\eea 
In the last normal ordered form, we have dropped a constant piece. The 
dimensionless \cc suggested by (\ref{LNO}) we denote simply by $f$.
\bdm
     f = \frac{\lambda}{\mu^2}    \label{f}
\edm
Since the normal order prescription in (\ref{LNO}) is relative to the 
vacuum of the symmetric phase theory, we may investigate the possibility of a 
critical value of the \cc $f$. Using (\ref{f}), the first line of (\ref{LNO}) 
may be rewritten as
\bdm
  {\cal L}_E = \frac{1}{2}(\nabla\phi)^2 +\frac{1}{2}\mu^2(1-3 f A_{\mu^2})
  \phi^2 + \frac{f\;\mu^2}{4} \phi^4   \label{LNO2}
\edm
On the lattice (fixed $a >0$), $A_{\mu^2}$ is finite; and we can argue that 
for small enough $f$, the exact effective potential is well-approximated by
the classical effective potential with its single minimum at $\phi_{cl}=0$.
For large $f$, the coefficient of $\phi^2$ in (\ref{LNO2}) becomes negative,
suggesting a transition to the broken symmetry phase. 
However the argument falls short at 
this point because for strong coupling one can not argue that the effective 
potential is well approximated by its tree level form. The argument was 
completed by Chang \cite{chang} by constructing a duality transformation 
from the strong coupling regime of (\ref{LNO}) to a weakly coupled theory 
normal ordered with respect to the vacuum of the broken symmetry phase.

There are several attempts in the literature (see Table 2) to compute 
the critical value, $f_c$, by various analytic approximations, with a rather 
large spread of answers. In this paper we report an accurate
numerical value by Monte Carlo simulation. The first step is to obtain 
the critical line in the $\bm2l,\laml$ plane. This determines $\bm2l(\laml)_
{crit}$ (see Fig.1). Recall that these values are  infinite volume 
extrapolations of finite volume Monte Carlo data.
 Then, combining
 (\ref{dmu}),(\ref{mrc3}) we obtain
\bdm
    \mu_{0_{\ell}}^2 = \mu_{\ell}^2 - 3\;\laml\;A_{\mu^2}  \label{mu2}
\edm 
In the infinite volume limit $A_{\mu^2}$ (\ref{A2}) has the integral 
representation
\bdm
    A_{\mu^2}= \int_0^{\infty}\;dt\;\exp(-\mu_{\ell}^2\;t)(\exp(-2t)I_0(2t))^2 
           \label{A3}
\edm
For any point away from the origin, (\ref{mu2}),(\ref{A3}) can be solved 
 numerically to determine $\mu_{\ell}^2(\laml)_{\mbox{crit}}$.
This is then extrapolated into the origin to determine
\bdm
           f_c = \lim_{\lambda_{\ell},\mu_{\ell}^2\rightarrow 0}  \label{fc}                               \frac{\laml}{\mu_{\ell}^2} |_{crit}    \label{fcrit}
\edm 

\section{\bf Simulations}
    The Monte Carlo simulation is based on the lattice action which 
regularizes the continuum theory (\ref{Leuc}).
\bdm
  {\cal A}= \sum_{\vec{n}}\left\{\frac{1}{2}\sum_{\nu=1}^{d}(\varphi(\vec{n}+
           \vec{e}_{\nu})-\varphi(\vec{n}))^2 \;+\frac{1}{2}\mu_{0_{\ell}}^2\;
           \varphi(\vec{n})^2 +\frac{\lambda_{\ell}}{4}\varphi(\vec{n})^4           
           \right\}
       \label{latA} 
\edm 
Periodic boundary conditions were imposed on $N \times N$ square lattices 
with $N = 32, 64, 128, 256$, and $512$. To reduce critical slowing down, our 
updating algorithm 
consisted of a standard Metropolis update (i.e. with a symmetric transition
matrix) alternating with a cluster algorithm updating the embedded Ising 
model.  The procedure is similar to that of Brower and Tamayo \cite{BT},
but we substitute a Wolff-type single-cluster algorithm for the Swendsen-Wang
multiple-cluster algorithm used there.  

Measurement of lattice quantities was performed every ten Metropolis+cluster 
cycles.  Each data collection run consisted of $10^4$ to $10^5$ measurements,
after an initial thermalization of at least $10^4$ cycles.  To assess the
effective number of statistically independent measurements, the integrated
autocorrelation time ($\ia$) was calculated for each run,  
\bdm
\ia= {1 \over 2} + \sum_{i=1}^{M}{s(i) \over s(0)}
\label{tauint}
\edm
where $s(i)$ is the autocorrelation separated by $i$ measurements and
$M$ is some number of measurement such that $s(M)$ is essentially noise. 
The largest
$\ia$ measured in our simulations was 8 measurements (80 update cycles)
and was typically much smaller.  Thus in every case the thermalization 
time exceeded $100 \ia$, so we expect that our lattices were well-thermalized
before we began collecting data.  We also tried to measure the exponential
autocorrelation time $\tau_{\rm{EXP}}$
from the first few (time) autocorrelation functions,
but these didn't appear to fall off as a single exponential.  As expected, 
however, the measured values were slightly smaller than the corresponding
$\ia$ for that run.  As a result of the small $\ia$ and large number 
of measurements, statistical
errors are typically quite small, generally smaller than the systematic
errors in the determination of the critical line in the $\lambda_L, \mu^2_L$
plane for finite volumes and in the extrapolation to the infinite-volume limit.

For each size lattice, we did looked at $\laml = 1.0, 0.7, 0.5, 0.25, 0.1$ and 
$0.05$, and for each $\laml$ scanned in  
$\bm2l$, starting in the symmetric phase 
and ending in the broken symmetry phase.  
We used two 
diagnostics to determine the critical value of $\bm2l$. The first was
the value of $\bm2l$ which produced
the maximum of the variance of the action (specific heat), which should 
diverge as $C_H \sim \ln{| \bm2l - \mu_{0_\ell c}^2|}$ as lattice size 
$L \rightarrow \infty$ \cite{TC}.
The second 
was based on the shape of the histogram of the distribution of $<\varphi>$. 
In the symmetric phase with $L \gg \xi$, the spatial correlation 
length, the probability distribution of $<\varphi>$ should be a 
single peak centered about zero, while in the broken-symmetry phase with 
$L \gg \xi$ the distribution should consist of two identical peaks 
at equal distance from $<\varphi>=0$.  For a fixed 
value of $\bm2l$, histograms of $<\varphi>$ will approach one of these 
distributions for $L$ sufficiently large.  To quantify the bimodality of
the distribution, we define 
$h$ to be the ratio of the number in the central bin to the largest 
number in any outlying bin. For a histogram which is a single peak 
centered about zero, $h > 1$. For a histogram which is two peaked, 
around $+<|\varphi|>$ and $-<|\varphi|>$, $h << 1$. 
The diagnostic for 
a given $\bm2l$ is the behavior of $h$ as the lattice size is increased.
For $\bm2l$'s which lead to symmetric phase in the infinite volume limit, 
$h$ increases with $L$ until it exceeds one. For $\bm2l$'s which 
lead to broken symmetry phase in the 
infinite volume limit, $h$ is rapidly decreasing with increasing $L$. 
For a narrow range of $\bm2l$ around the infinite volume critical 
value, this behavior may not stand out until one gets to quite large 
lattices. In practice, on the lattices we used, from these observations 
we obtained upper and lower bounds on the infinite volume critical 
value of $\bm2l$ which are closer togther than the estimated statistical 
and systematic error in the value obtained from extrapolation of the 
position of the peak of the variance of the action. (In all cases, 
within that estimated error the extrapolated value is consistent with the 
bounds). 

\section{Analysis}

In the second column of Table 1 we give our estimates of the critical value of
$\bm2l$ for each of the values of $\laml$ listed above, extrapolated to the infinite volume limit. 
(They are the 
input for Fig. 1). They may be compared with the values found by Toral 
and Chakrabarti \cite{TC} seven years ago, extrapolating from much smaller 
lattices. Within the estimated errors they are consistent, except for 
the smallest $\laml$, which is closest to the continuum limit and requires 
larger lattices to accomodate both large $L$ and small $a$.  
In the third column of Table 1 we give the corresponding critical 
values of $\mu^2$ as determined from (\ref{mu2}), (\ref{A3}), and in 
the fourth column we give the corresponding values of $\lambda \over \mu^2$.

\begin{table}
\begin{center}
\begin{tabular}{|c|c|c|c|}
$\lambda$ & $\mu_{0C}^2$ & $\mu^2_C$ & ${\lambda \over \mu^2}$ \\[0.1in]
\hline 
1.0  & -1.270(3)   & 0.0980(8)     & 10.204(80)  \\
0.7  & -0.9516(8)  & 0.06844(23)   & 10.228(33)  \\
0.5  & -0.7210(10) & 0.0489(3)     & 10.225(63)  \\
0.25 & -0.4035(5)  & 0.0242(2)     & 10.33(8)    \\
0.10 & -0.1878(5)  & 0.009615(140) & 10.40(15)   \\
0.05 & -0.0998(3)   & 0.00492(9)     & 10.16(19)
\end{tabular}
\caption{Determination of the Phase Transition Line for Different $\lambda_l$}
\end{center}
\end{table}
In order to get a feel for the systematic 
errors we have done the extrapolation to the continuum limit 
in a number of different ways.  We have 
plotted $\mu^2$ vs$\lambda$ and taken the inverse of the slope at the origin to
determine the critical value of $f$.  
This is shown in Fig. 3.  
\begin{figure}
\centering
\epsfysize=2in  
\hspace*{0in}
\epsffile{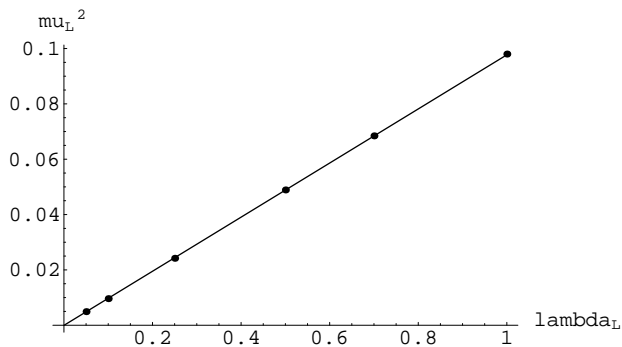} 
\caption{Plot of Phase Transition Line in the $\mu^2_l, \lambda_l$ Plane}
\label{mu2l_vs_lambdal} 
\end{figure}
Note that 
the data fall very well on a straight line ($\chi^2=0.65$) which passes 
throught the origin within the estimated error.  As a small variation on this
we have redone the fit forcing the line to go exactly through the origin.  The 
two values of $f_c$ determined in this manner are 10.23(3) 
and 10.24(2) respectively.
Alternately, we have extrapolated the values of $f={\lambda \over \mu^2}$ to
the continuum limit ($\lambda_{lat} \rightarrow 0$) as shown in Fig. 4.
We assume that we approach the continuum limit along 
a `line of constant physics',
i.e. once we are close enough to critical ($\xi_{lat} \gg 1$), dimensionless 
physical quantities should not change their values as the lattice spacing is 
reduced. Thus we fit the data in Fig.\ref{f_vs_lambdal} with a horizontal line. 
The $\chi^2 = 0.59$ is consistent with this assumption and the result for 
$f_c$ is 10.24(3).

Although the errors originated as statistical errors in simulations on finite 
lattices
the subsequent extrapolations have introduced systematic errors larger than the
statistical errors.  Thus the assignment of the final result and errors 
are just 
based on the consistency of the above numbers. We conclude that the critical 
value of $\frac{\lambda}{\mu^2}$ is 10.24(3).
\begin{figure}
\centering
\epsfysize=2in  
\hspace*{0in}
\epsffile{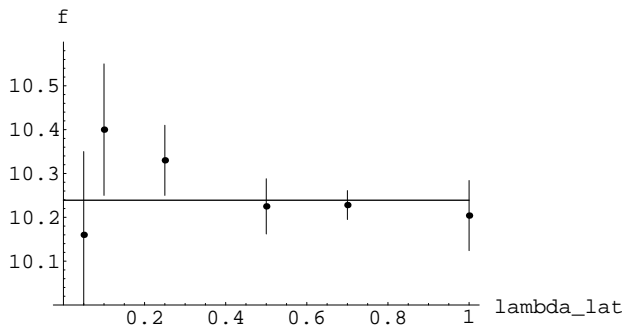} 
\caption{Extrapolation to determine the critical value of 
$f$ }  
\label{f_vs_lambdal} 
\end{figure}

\section{Analytic Approximations}

The above result may be compared with various approximate analytic 
calculations.
The simplest approach would be to consider the one-loop effective potential.
In $d = 2$ this is
\bdm
V_{eff}= {\mu^2 \over 2} \phi^2 + {\lambda \over 4} \phi^4 -{ {\mu^2 + 3 \lambda \phi^2}
\over {8 \pi}} ( \ln{{\mu^2 + 3 \lambda \phi^2} \over \kappa^2} -1 )
\edm  
With the assignment $\kappa^2=\mu^2$ this $V_{eff}$ gives a 
first-order phase transition
for $f_c=6.6$.

A compendium of nonperturbative analytic approximate calculations 
has been compiled 
in Ref. \cite{hauser}.  We include these results for purposes of 
comparison with the numerical MC simulation result.  

\begin{table}
\begin{center}
\renewcommand{\arraystretch}{1.2}
{
\begin{tabular}{c|c|c|}
Approximation & Result & Reference  \\
\hline
Non-Gaussian Variational & 6.88 & \cite{polley} \\
Discretized Light-Front & 7.316,5.500 & \cite{harind1},\cite{harind2} \\
Coupled Cluster Expansion & $3.80 < f_c < 8.60$ & \cite{funke} \\
Connected Green Function & 9.784 & \cite{hauser} \\
Gaussian Effective Potential & 10.211 & \cite{chang} \\
 & 10.272 & \cite{hauser}  
\end{tabular}
}
\caption{Analytic Approximations of the Critical Value of $f$}
\end{center}
\end{table}

There is also the issue of the order of the phase transition.  According to the
Simon-Griffiths theorem \cite{Simon}, 
the phase transition is second order.  In the analytic
approximations the order of the phase transition is determined; the one-loop
effective potential and the Gaussian effective potential predict a 
first-order phase transition, while the other correctly predict a second-order 
phase transition.  In our numerical calculation of the of the critical 
couping constant we have made no effort to distinguish between weakly first-order
and second-order phase transitions.

\section{Conclusions}

We have calculated an accurate numerical value of the critical coupling
constant using Monte Carlo simulation. With this we can evaluate the accuracy
of analytic approximation methods.  It is interesting to observe that the Gaussian
effective potential result for the critical coupling 
is consistent with the accurate numerical result, although it gives incorrectly the 
order of the phase transition.   

\section{Acknowledgements}

We would like to thank Prof. R. Mawhinney for writing some of the programs
used in this project.  W. Loinaz was supported in part by the 
U.S. Department of Energy under the grant \#DE--FG05--92ER40709--A005. 
R. Willey thanks Prof. Carlos 
Aragao de Carvalho for hospitality at CBPF Brazil where this work was begun.

\end{document}